# EXCERTOS DA HISTÓRIA KEYNESIANA.

**Gustavo Lima Moura[1]**

[1]Mestrando em Economia Política Mundial – Universidade Federal do ABC (UFABC) – São Bernardo do Campo – SP – Brasil.

`moura.g@ufabc.edu.br`

***Abstract.*** *This article analytically describes the contributions of Keynesian theory in the post-World War I context, by means of a synoptic reading of the bibliographical sources indicated, the Keynesian theory within its own historical and philosophical context. The discussion covers the main concepts of his theory. The aim is, through the excerpts highlighted in the article, to briefly contextualize the origins of the ideas of John Maynard Keynes' economic thought.*

***Resumo.*** *Este artigo descreve analiticamente as contribuições da teoria keynesiana no contexto pós-Primeira Guerra Mundial, por meio de uma leitura sintópica das fontes bibliográficas indicadas, a teoria keynesiana dentro de seu próprio contexto histórico e filosófico. A discussão abrange os principais conceitos da sua teoria. O objetivo é através dos excertos destacados pelo artigo, contextualizar brevemente as origens das ideias do pensamento econômico de John Maynard Keynes.*

## 1. A GRANDE GUERRA.

Dois aspectos principais ajudam a compreender o embrião da teoria keynesiana, especialmente em relação à moeda: o fim da era vitoriana, marcada pela estabilidade econômica e pelo *laissez-faire*, e a perda do papel social da poupança burguesa.

Durante a era vitoriana, o crescimento econômico era impulsionado pela poupança dos capitalistas, que acumulavam recursos e os reinvestiam, promovendo a expansão da economia. Esse agente poupador conferia legitimidade ao capitalismo clássico (*laissez-faire*), pois parte do que obtinha retornava à produção, beneficiando o conjunto da sociedade. Keynes argumentava (*The Economic Consequences of the*

*Peace*) que a paz de 1919 impunha a necessidade de uma nova abordagem econômica. As teorias e políticas vigentes até então precisavam ser reconsideradas, pois o cenário pós-guerra demandava respostas diferentes para desafios inéditos.

Com o fim da guerra, a poupança burguesa já não desempenhava mais esse papel central. Tornou-se um fenômeno disperso devido às transformações estruturais na economia global, como a destruição de capital, a desorganização das finanças públicas e a necessidade de reconstrução dos países europeus. O capital já não estava exclusivamente nas mãos de grandes empresários e proprietários, mas se diluía em novas formas de acumulação, como depósitos bancários, pequenos investimentos e títulos de guerra adquiridos pela população durante o conflito. Nesse contexto, o capitalismo tradicional do *laissez-faire* foi desafiado, uma vez que a poupança deixou de cumprir a função social de reinvestimento na produção, fundamental para a estabilidade econômica da era vitoriana, encerrada em 1901.

Nesse sentido, o debate Keynesiano com a era clássica da economia política, não se trata apenas da modelagem no contexto de crise, espacial e temporalmente localizados, como apresentada pela realidade vigente no período do pós-primeira guerra mundial, seu modelo e sua interpretação, como apresentado em outras obras (ex.: *A Treatise on Money* (1930); *Treatise of Probability* (1921)[1], etc.), estão convergindo para a análise do novo capitalismo, instável, e sua reformulação à causas do câmbio da distribuição da produção e do emprego.

As causas reais, monetárias e financeiras, de desequilíbrio econômico, para Keynes, demonstra seu afastamento das concepções clássicas de autorregulação e equilíbrio da lei de Say; para ele, a incerteza inerente ao dito período, gera um comportamento receoso dos agentes em relação à liquidez e ao investimento, o que implica em um baixo investimento derivado da poupança e consequentemente em um desequilíbrio nos salários nominais e no nível de emprego, o que contradiz as teorias clássicas.

Com isso, a poupança, deixa de ter o aspecto exclusivamente benéfico para o crescimento econômico, e começa a gerar um descompasso no âmbito macroeconômico, o que Keynes chama de "paradoxo da parcimônia"; isso é, um aumento generalizado

---

1 Obra vital para compreender o conceito do "*Animal Spirits*", discorreremos mais afrente no artigo.

nas poupanças, reduzem a demanda efetiva freando a produção e o emprego.

Essa nova realidade, mostrava a Keynes que o desemprego poderia não ser necessariamente um fenômeno sazonal, como previam as teorias clássicas, e sim algo prolongado. E para Keynes, a resposta a esse fato, era política, ele, em primeiro lugar defendia uma política monetária de intervenção, com baixas taxas de juros à fim de baratear a obtenção de moeda e consequentemente estimular a produção e o consumo; em segundo lugar obras públicas, pois como a demanda global havia caído, o Estado necessariamente precisa gerar um mecanismo compensador e em terceiro lugar, o protecionismo moderado, pois redireciona o consumo para o mercado interno ao invés de estimular o consumo pela importação.

## 2. A TEORIA GERAL

A obra "*The General Theory of Employment, Interest and Money*" publicada em 1936, centra aquilo que podemos chamar de revolução keynesiana, porque seu objetivo é rechaçar toda a construção teórica do liberalismo clássico, criando uma nova explicação para o desemprego, nesse novo período do capitalismo, onde os altos salários artificialmente elevados por conta dos sindicatos ou outras variáveis que impediriam o ajuste natural entre oferta e demanda de trabalho, não são a causa do desemprego, como prediziam os economistas clássicos.

O argumento do livro se concentra na questão do desemprego estrutural, e não meramente conjuntural. Segundo o autor, se um mercado ajusta as quantidades de emprego, a demanda agregada global não se comportará como prevê a teoria clássica. Isso ocorre porque, à medida que o nível de emprego aumenta, a produção também cresce. No curto prazo, dada a capacidade produtiva existente, os empresários determinarão a produção com base nos lucros necessários para cobrir seus custos e não o contrário, como afirmava Say. Em outras palavras, a oferta não cria a própria demanda.

Keynes, por sua vez, argumenta que é o nível de emprego e os salários (ou seja, a liquidez) que determinam a produção. Esse é o princípio central da teoria da demanda efetiva, expressa na equação $Y = C + I$. Nesse sentido, as variáveis determinam o nível de atividade econômica.

A desestabilização na curva de Say, para Keynes, é causada pela ruptura com o mundo clássico, onde a incerteza inerente às decisões econômicas e o comportamento dos agentes em relação ao investimento e liquidez, frenam o processo de retorno das poupanças ao meio produtivo causando a instabilidade na inversão, a demanda agregada (total de gastos), se torna o elemento central para determinar os níveis de emprego e produção, pois quando ocorre uma insuficiência na demanda agregada, em virtude dos agentes poupadores não investirem face à incerteza do futuro, há um aumento do desemprego involuntário.

## 3. A POLÊMICA TEORIA DA PROBABILIDADE LÓGICA, FRANK RAMSEY.

Uma polêmica interessante na história do pensamento keynesiano, se encontra na teoria da probabilidade lógica, derivada da obra "*Treatise of Probability*" (1921), onde Keynes, muda sua opinião sobre a teoria da probabilidade lógica não frequencialista – que se aproxima da filosofia do racionalismo e do a priori – ante as críticas de um matemático chamado Frank Ramsey, no seu artigo "*Truth and Probability*" (1926).

> (...) O Sr. Keynes acredita que a relação de probabilidade não numérica corresponde a um grau não numérico de crença racional, mas que os graus de crença, que eram sempre numéricos, não correspondiam um a um com as relações de probabilidade que os justificavam. Pois é concebível, suponho, que os graus de crença possam ser medidos por um psicogalvanômetro ou algum instrumento semelhante, e o Sr. Keynes dificilmente desejaria que as relações de probabilidade pudessem ser todas medidas derivativamente com as medidas das crenças que elas justificam. (RAMSEY, p. 9, 1926, tradução nossa).

O ponto no que toca Ramsey é a dificuldade na medição das crenças, ou a possibilidade de correlação entre crenças e probabilidade. Ramsey, descreve de uma forma bastante cética a não percepção das relações de probabilidade e as crenças psicológicas dos agentes econômicos, não sendo possível afirmar o que Keynes descreve no seu "*Treatise on Probability*"; por falta de evidência empírica direta na relação de um a um entre graus de crença e probabilidades. Criticando a crença de Keynes sobre a natureza da necessidade lógica, seguindo o "*Tractatus Logico-Philosophicus*" (1921) de Wittgenstein[2].

> "A proposição constrói o mundo com a ajuda de andaimes lógicos, e por isso é possível, na proposição, também se ver, *caso* ela for verdadeira, como tudo que é lógico está. Pode-se de uma proposição falsa *tirar conclusões*." (WITTGENSTEIN, p. 169, 192).

Onde a principal divergência com a teoria da probabilidade Keynesiana, se dá no campo da natureza da probabilidade e das proposições lógicas, pois enquanto Keynes na obra via a natureza como algo racional, baseado na lógica entre as proposições, Wittgenstein e Ramsey, percebiam a lógica como algo voltado para tautologias e não crenças.

> A probabilidade é, vide Capítulo 11 (§12), relativa em um sentido aos princípios da razão humana. O grau de probabilidade que é racional para nós entreter não pressupõe perfeita compreensão lógica, e é relativo em parte às proposições secundárias que de fato conhecemos; e não depende de se uma compreensão lógica mais perfeita é ou não concebível. É o grau de probabilidade ao qual esses processos lógicos levam, dos quais nossas mentes são capazes; ou, na linguagem do Capítulo II, que essas proposições secundárias justificam, que de fato conhecemos. Se não tomarmos esta visão da probabilidade, se não a limitarmos desta forma e a tornarmos, até certo ponto, relativa aos poderes humanos, estaremos completamente à deriva no desconhecido; pois nunca poderemos saber que grau de probabilidade seria justificado pela percepção de relações lógicas que somos, e devemos sempre ser, incapazes de compreender. (Keynes, 1921, p. 164, apud RAMSEY, 1926, p. 11, tradução nossa).

Sendo assim, Keynes responde a Ramsey com uma teoria que reformularia a forma de comportamento do investimento substancialmente a partir do argumento de que os investimentos se baseavam nas expectativas de futuro, que são incertas, mas que se convertem em convenções grupais, e é a partir da interação entre os investidores, onde forma-se uma expectativa futura, uma atmosfera de pessimismo ou otimismo econômico no mundo dos investimentos. E isso, Keynes, à resposta do artigo de Ramsey ao seu tratado sobre probabilidades, chama de *Animal Spirits.*

---

2 Filósofo da mente, linguagem, matemática e lógica.

> Mesmo para além da instabilidade devida à especulação, existe a instabilidade devida à caraterística da natureza humana de que uma grande parte das nossas atividades positivas depende do otimismo espontâneo e não de uma expetativa matemática, seja ela moral, hedonista ou econômica; a maior parte das nossas decisões de fazer algo positivo, cujas consequências totais se prolongarão por muitos dias, só podem ser tomadas em resultado dos *animal spirits* - de um impulso espontâneo para a ação em vez da inação, e não como resultado de uma média ponderada de benefícios quantitativos multiplicados por probabilidades quantitativas. (Keynes, 1936 ed. 2017, p. 139, tradução nossa).

O conceito de *Animal Spirits* surge especificamente da obra de Descartes (*Traité de l'Homme* – 1664), e se consolida na filosofia de David Hume (*Treatise of Human Nature* – 1739), ao descrever os instintos e paixões que governavam a natureza humana na contramão da racionalidade pura (ou razão).

A resposta Keynesiana, que engloba os *Animal Spirits,* se dá posteriormente à crítica de Ramsey, já no livro *The General Theory of Employment, Interest and Money* (1936)*,* quando ele busca entender a natureza da ação humana em virtude do desconhecimento futuro, com isso, o que Keynes entende é que quando há o desconhecimento futuro, mas a necessidade de uma 'atividade positiva', os homens atuam mediante convenções grupais em torno a como será o futuro, sob a ideia clara, de que o impulso espontâneo, tal qual ele descreve, não é medido quantitativamente, e portanto, está sujeito a instabilidade, pois qualquer pequeno dado futuro, pode tornar as efêmeras convicções sobre o futuro, demasiado frágeis. Por isso o investimento é a variável instável do capitalismo.

## **4.** *HOW TO PAY FOR THE WAR* **(1940)**

A obra *How to pay for the war* (1940), é uma obra realizada em um momento extemporâneo quando se acerca o período da segunda-guerra mundial, quando ele é nomeado assessor do ministério do tesouro durante todo o período da guerra, o qual se trata da compilação de vários artigos onde Keynes levanta a tese de que os níveis de emprego britânicos haviam sido recuperados pelo esforço bélico, e em tal ponto se dá o aquecimento, que está resultando no aumento dos preços.

Porém, uma economia com inflação, teria sérios problemas de aprovisionamento e de pagamentos, câmbios, etc. Ao que Keynes enxerga uma mudança na forma de pagamento da guerra, isso é, ao mesmo tempo que a guerra gerou o crescimento econômico e aumentou os níveis de liquidez, acabando com a depressão internacional; ela causa uma alteração de equilíbrio, o problema levantado na questão de como pagar a guerra, não é mais o desemprego como muito bem ressaltado na "*economic consequences of the peace* (1919)", e sim a inflação.

Nessa obra, a solução proposta por Keynes, era exatamente o que não foi feito na primeira guerra mundial, ele diz em essência ser preciso tributar as rendas, arrecadar esses impostos – que reduziriam a capacidade de compra dos trabalhadores (embora também fossem estabelecidos mínimos isentos) – e depositá-los em uma caderneta postal, na caixa de poupança postal. E então, esses fundos seriam liberados ao fim da guerra, pois, quando a guerra terminasse, viria a depressão, e essa capacidade de compra tornaria a transição muito mais pacífica.

E por outro lado, ele também defendia nessa série de artigos compilados, que a economia do pós-guerra, se refere necessariamente ao fim do padrão ouro, com a criação de um fundo de compensação de dívidas, obrigando os países com excedentes a financiarem os países com deficit na sua conta corrente; ao que resulta, posteriormente na vitória da posição americana como é bem conhecido, com o acordo de Bretton Woods em 1944, onde foi fundado do Fundo Monetário Internacional e o Banco de Reconstrução e Fomento, depois conhecido como Banco Mundial, nos moldes da proposição do economista e funcionário do tesouro americano Harry White.

## 6. LEGADO

A difusão teórica keynesiana, foi amplamente exitosa em todas as revistas científicas do mundo naquele período, diversas revistas começaram publicar artigos sobre Keynes. E isso, por duas razões, primeiro porque a revolução Keynesiana significou uma provocação frente a teoria clássica, onde o modelo IS-LM representava a teoria geral, frente a teoria clássica da lei de Say, o que gera uma grande altercação entre os acadêmicos do período, que argumentavam seus artigos.

O que gerou uma progressiva mudança na forma de fazer economia política, especialmente por influência de um conjunto intermediário de autores que estavam estabelecendo uma modelagem matemática da teoria geral, que era uma revisão da teoria geral, e que ao final, gera a síntese neoclássica-keynesiana.

A síntese neoclássica-keynesiana, a partir de 1937 com John Hicks e posteriormente na década de 1950 com Paul Samuelson, toma como premissa elementos de economia política clássica – muito influenciado indiretamente por Walras, e concilia com a revolução keynesiana, criando um modelo de **equilíbrio geral**. Basicamente se distinguindo dois mercados, um no qual o equilíbrio se dá entre poupança e investimento que dará lugar a curva IS e o outro oferta e demanda de dinheiro no mercado monetário, e a sua vez, o equilíbrio entre ambos, pode ser ou não, de plena ocupação, e a discussão que se propõe na síntese neoclássica-keynesiana, é se o modelo IS-LM de Keynes realmente era um modelo geral.

Porém, à discordância da sintese neoclássica-keynesiana, Keynes não havia proposto exatamente o que pressupõe a interpretação mais moderna, pois a sua teoria geral se aplicava, em situações de rigidez de salários ou o mercado monetária tivesse dificuldades de ajuste (armadilha de liquidez), poderia ser haver problemas em encontrar um equilíbrio na plena ocupação.

De qualquer forma, o modelo IS-LM foi desenvolvido econometricamente, se criaram departamentos de pesquisa para fazer estimações da contabilidade nacional (tal qual no período entre guerras com a ascensão do institucionalismo), e pouco a pouco, foi-se popularizando dentro do campo de pesquisa econômica, a predominância de que o modelo keynesiano explicava macroeconomicamente o sistema econômico e ademais dava soluções político-econômicas consistentes.

Por fim, a popularização do paradigma keyenesiano representou uma resposta científica à grande depressão e segunda guerra mundial, oferecendo um *framework* importante para os Estados intervirem ativamente na regulação econômica e criando um novo tipo de ciência social aplicada, que perdura até os dias atuais, através das suas adaptações e distintas correntes.

**Referências**

estudo introdutório de Luiz Henrique Lopes dos Santos. Introdução de Bertrand Russell. 3. ed. São Paulo: Editora da Universidade de São Paulo, 2001.